\documentclass[twocolumn,showpacs,preprintnumbers,amsmath,amssymb]{revtex4}
\usepackage[colorlinks=true,urlcolor=blue,allcolors=blue]{hyperref}
\usepackage{graphicx}
\usepackage{dcolumn}
\usepackage{bm}

\begin{document}
\title{Topologically ordered zigzag nanoribbon:  $e/2$ fractional edge charge,  spin--charge separation, and ground--state  degeneracy}
\author{ S.-R. Eric Yang$^{1}$\footnote{corresponding author, eyang812@gmail.com},  Min-Chul Cha$^{2}$,  Hye Jeong  Lee$^{1}$, and Young Heon Kim$^{1}$}
\affiliation{ $^{1}$Department of Physics, Korea University, Seoul, Korea\\ 
$^{2}$Department of Photonics and Nanoelectronics, Hanyang University, Ansan, Korea\\}

\begin{abstract}
We numerically compute the density of states (DOS) of interacting disordered zigzag graphene nanoribbon (ZGNR) having midgap states showing $e/2$ fractional edge charges. The computed Hartree--Fock DOS is linear at the critical disorder strength where the gap vanishes.
This implies an $I\mbox{-}V$ curve of $I\propto V^2$. Thus, $I\mbox{-}V$ curve measurement may yield evidence of fractional charges in interacting disordered ZGNR.  We show that even a weak disorder potential acts as a singular perturbation on zigzag edge electronic states, producing drastic changes in the energy spectrum.   Spin--charge separation and fractional charges  play a key role in the reconstruction of edge antiferromagnetism.     Our results show that an interacting disordered ZGNR is a topologically ordered Mott-Anderson insulator.

\end{abstract}
\maketitle

\section{Introduction}

\begin {figure}[!hbpt]
\begin{center}
\includegraphics[width=0.5\textwidth]{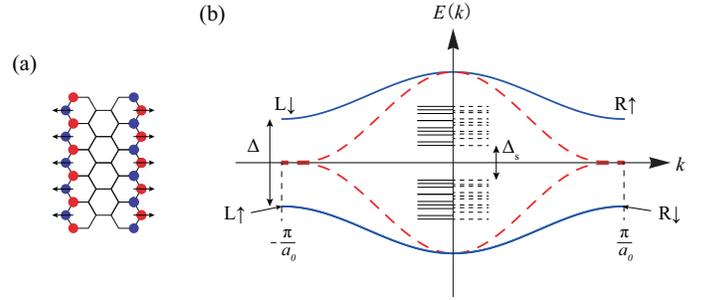}
\caption{(a) Zigzag edge antiferromagnetism of interacting ZGNR without disorder.    (b)   Schematic of interacting (solid curves) and noninteracting  (dashed curves)  ZGNR band structures. The unoccupied and occupied  
states near the wave vectors $k=\pm \pi/a_0$ are shown: $R$ and $L$ represent states confined on the right and left zigzag edges, respectively ($a_0=1.73a$  is the unit cell length of the ZGNR and $a=1.42 \AA$   is the C-C distance).  
The small arrows indicate spins.  The spin-split energy levels of the spin-up (solid lines) and spin-down (dashed lines) {\it gap--edge} states of the interacting disordered ZGNRs are shown. These states decay exponentially from the zigzag edges.  In the limit of   an infinitely long ribbon the gap   $\Delta_s$ may vanish and a soft gap can develop.  {\it Another degenerate ground state can be obtained by exchanging $\uparrow$ 
and $\downarrow$ spins}.}
\label{degGap}
\end{center}
\end{figure}

\begin {figure}[!hbpt]
\begin{center}
\includegraphics[width=0.3\textwidth]{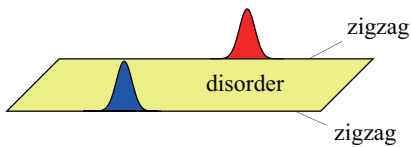}
\caption{Two $e/2$ fractional zigzag edge charges of an interacting disordered ZGNR.  Note that red (blue) probability  density means that the wave function has A (B) chirality, i.e., it is finite only on   A (B) carbon atoms. Since there is negligible tunneling between these sites we will call this type of state a mixed chiral state.}
\label{fracchar}
\end{center}
\end{figure}

Graphene has numerous remarkable properties \cite{Nov,Zhang,Neto}. One prominent feature is that, in the absence of disorder, zigzag graphene nanoribbons (ZGNRs) can support  chiral symmetry protected topological (SPT) \cite{compoly,Wen1,Wen2,Ryu,Jeong1,Jeong0} edge states displaying an integer charge \cite{Fujita}. Disorder has profound effects on ZGNRs. In particular, an interacting disordered ZGNR becomes a Mott--Anderson insulator \cite{dimension,Insul,Dob} with spin-split energy levels \cite{Jeong2}. 
Furthermore, localized gap-edge states reduce the size of the gap between the occupied and unoccupied midgap states with energies $-\Delta_s/2$ and $\Delta_s/2$, respectively, to $\Delta_s$ (see Fig.~\ref{degGap}).
In the weak disorder regime, solitonic midgap states \cite{Su,yang1} may have an $e/2$ fractional charge on each of the opposite zigzag edges, i.e., {\it there is one for each edge} \cite{Jeong2}, where $e$ is the electron charge, see Fig.~\ref{fracchar}. These  fractional charges have small {\it disorder}-induced charge variances. In addition, the charge fractionalization is protected against {\it quantum} charge fluctuations by the nonzero $\Delta_s$. 
Here, $\Delta_s\lesssim 10^{-2}\Delta\sim 1\,\textrm{THz}$, where $\Delta$ is the gap value; this is sufficiently large that quantum charge fluctuations can be ignored (see Girvin \cite{Girvin}). In the absence of disorder, typically, $\Delta\sim 0.2t$ \cite{Yang}, where $t\sim 3\,\text{eV}$ is the hopping constant.  

\begin {figure}[!hbpt]
\begin{center}
\includegraphics[width=0.35\textwidth]{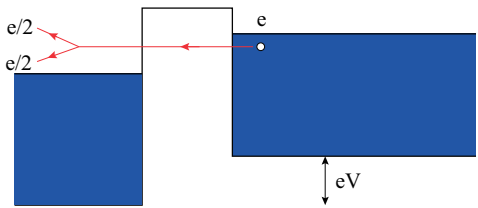}
\caption{A tunneling electron is fractionalized when it enters an interacting disordered ZGNR. }
\label{tunniv}
\end{center}
\end{figure}

An excellent opportunity to observe these boundary charges has recently arisen, as rapid progress has been made in the fabrication of atomically precise GNRs \cite{Cai2}.
The chiral Luttinger liquid theory of fractional quantum Hall edges \cite{Wen,Cha} predicts an $I\mbox{-}V$ curve of $I\propto V^{1/\nu}$. The corresponding DOS is given by 
\begin{equation}
\rho(E)\propto E^{1/\nu-1}, 
\end{equation}
where $\nu$ is the filling factor, and the energy $E$ is determined from the Fermi energy (these edges support gapless excitations). This predicted $I\mbox{-}V$ curve has been experimentally confirmed \cite{Gray}. 
It should be noted that Laughlin quasiparticles have an odd denominator fractional charge $e\nu$, and an even denominator fractional charge $e/2$ 
is not found in fractional quantum Hall systems. The aforementioned $I\mbox{-}V$ curve may be derived heuristically by assuming that a tunneling electron fractionalizes into $m=1/\nu$ fractionally charged quasiparticles \cite{GY}, where 
\begin{equation}
e\rightarrow e/m+\dots +e/m. 
\end{equation}
(This tunneling process is illustrated in Fig.~\ref{tunniv}).  However, the chiral Luttinger liquid theory does not apply to ZGNRs. Furthermore, the gap--edge states are all localized along the ribbon direction, in contrast to the fractional quantum Hall edge states. Moreover, the average edge charge of the gap--edge states with energies within a small interval $\delta E$ is $e/2$; however, significant disorder-induced charge fluctuations may occur. Some of these states are more localized on the left or right zigzag edges. This tendency increases as the electron energy deviates from  $\pm\Delta_s/2$. 
Despite this,
if we apply the above heuristic argument to a ZGNR with $m=2$, then we find that  the $I\mbox{-}V$ curve is given by
\begin{equation}
I\propto\int d\epsilon_1 \int d\epsilon_2 \theta (eV-\epsilon_1 -\epsilon_2)\propto V^2, 
\end{equation}
where
$\theta$ and $\epsilon_{1,2}$ are the step function and quasiparticle energies, respectively. This $I\mbox{-}V$ curve is equivalent to a {\it linear} tunneling DOS.     A topological insulator  is usually not significantly  affected by
a  disorder potential, but the SPT phase of a  ZGNR is
profoundly changed by disorder.  However,   the physical processes involved  in this  effect and  the properties of the  interacting  disordered state are not well understood.  

In this study, we propose an experiment that may provide evidence of the presence of $e/2$ fractional charges in interacting disordered ZGNRs. 
We compute the DOS of an interacting disordered ZGNR and find that, for the critical disorder strength where the ZGNR supports gapless excitations (i.e., where $\Delta_s$ vanishes), our computed Hartree--Fock (HF) DOS is linear near the Fermi energy. This finding is in agreement with the heuristic argument given above.
In addition, our results show that even a weak disorder potential behaves similar to a {\it singular} perturbation on zigzag edge electronic states, generating 
drastic changes in the energy spectrum.  It also induces a magnetic zigzag edge reconstruction in which
fractional edge charges and {\it spin-charge separation} play a significant role.  Moreover, disorder also changes  an SPT phase to  a {\it topologically ordered} phase \cite{Wen1, topoph}.

\section{Model}

\begin {figure}[!hbpt]
\begin{center}
\includegraphics[width=0.25\textwidth]{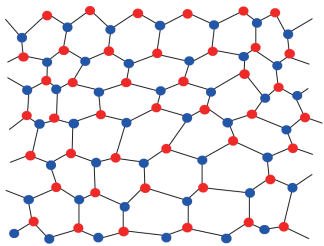}
\caption{Off--diagonal disorder: Random network of hexagons consisting of A and B carbon atoms.  A zigzag edge site is connected  to {\it two} other carbon atoms while  a site away from the edges is connected to {\it three} other carbon atoms.  The hopping parameter $t$ is  not  the same for all sites.  The zigzag edges have definite chirality,  consisting A or B carbon atoms.  In contrast, armchair edges have mixed chirality.}
\label{offdia}
\end{center}
\end{figure}

\begin {figure}[!hbpt]
\begin{center}
\includegraphics[width=0.2\textwidth]{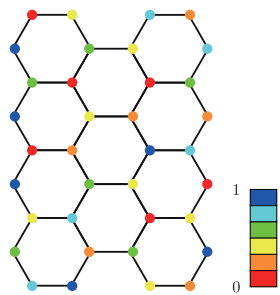}
\caption{ Diagonal disorder:  Site energies are varied randomly.  Colors represent strength of on--site disorder potential.  Again a zigzag edge site is connected  to two other carbon atoms while a site away from the edges is connected to three other carbon atoms.  But the hopping parameter $t$ is the same for all sites.}
\label{diadis}
\end{center}
\end{figure}

\begin {figure}[!hbpt]
\begin{center}
\includegraphics[width=0.35\textwidth]{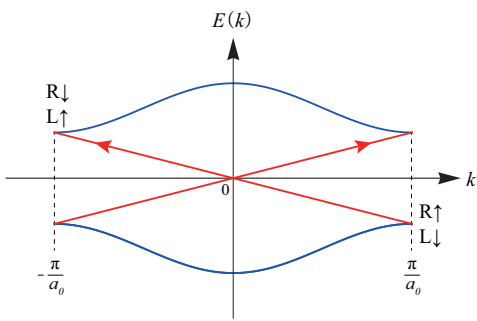}
\caption{ States localized on the right and left zigzag edges are represented, respectively, by $R$ and $L$. 
The long arrows indicate  the coupling, induced by a short-ranged disorder potential, between states $R \uparrow$ and $L\uparrow$   or  $R \downarrow$ and $L\downarrow$.
 }
\label{disorderscatt}
\end{center}
\end{figure}

\begin {figure}[!hbpt]
\begin{center}
\includegraphics[width=0.35\textwidth]{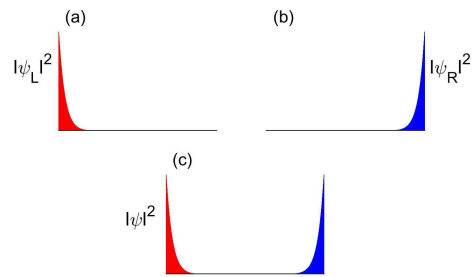}
\caption{   Schematic drawing of the site probability distribution $|\psi|^2$ of  two degenerate edge states with wave functions $\psi_L$ and $\psi_R$ is shown.   They are  localized on the (a) left and (b) right edges, respectively.  (c) Disorder couples these states and can generate antibonding    and bonding edge states with the wave functions  $\psi=\frac{1}{\sqrt{2}}(\phi_L-\phi_R) $ and $\psi=\frac{1}{\sqrt{2}}(\phi_L+\phi_R)$, respectively.}
\label{endsts}
\end{center}
\end{figure}

There are two types of disorder, namely  diagonal and off--diagonal disorder.   We model off--diagonal disorder by randomly varying  the nearest-neighbor hopping parameters,  see Fig.~\ref{offdia}.  However, since the results of off--diagonal disorder and diagonal disorder are similar 
we mainly report, in this study,   on the results of diagonal disorder, shown schematically in Fig.~\ref{diadis}.  In diagonal disorder    $N_{imp}$ defects or short--ranged impurities are randomly placed at carbon 
sites $\vec{R}_j$.  
Let us analyze the scattering of  left and right edge  states  by a short-ranged disorder potential.   
Consider
a spin-up electron at  $k=\frac{\pi}{a_0}$ with the wave function $\phi_{R\uparrow}$
localized on the right zigzag edge.  
For a short-ranged potential, 
a significant wave vector transfer  in a backscattering  occurs  for $|k - k'|\sim 1/a_0$  \cite{Lima}.  Such a short-ranged disorder potential   couples the {\it chiral}
zigzag edge state $R\uparrow$ to another {\it chiral} zigzag edge state $L\uparrow$ on the opposite zigzag edge at    $k= -\frac{\pi}{a_0}$, as shown in  Fig.~\ref{disorderscatt}  [their wave functions $\phi_{R}$   and  $\phi_{L}$ are depicted in Figs.~\ref{endsts}-(a) and \ref{endsts}(b)].
This process produces the bonding 
or antibonding  edge  state with the wave function $\frac{1}{\sqrt{2}} (\phi_L +\phi_R )$ or $\frac{1}{\sqrt{2}} (\phi_L -\phi_R )$.  
The probability density of one of these  states is shown schematically in Fig.~\ref{endsts} (c) (a mixed chiral state).  These states display charge fractionalization with $1/2$ charges on the left and right zigzag edges. But states 
with uneven fractions may also be generated.   Numerical calculations  are needed to determine the distribution of these edge charges.

The strength
of the  potential $\epsilon_j $ is chosen randomly from the energy interval $[-\Gamma,\Gamma]$.  The  values of $\epsilon_j $  depend on the type of  charged impurities in the substrate and defects in graphene.  
The number of impurities or defects is also relevant in determining the strength of the disorder potential.  In the Born approximation disorder strength is characterized by the parameter $\Gamma\sqrt{n_{imp}}$, where $n_{imp}=N_{imp}/N$ is the  ratio between the  numbers of impurities and total carbon atoms.   The range  of an impurity potential is important in determining properties of the Dirac electrons in ZGNRs \cite{Ando,Lima,LongShort}.   However, a  short-ranged disorder potential gives more robust charge fractionalization \cite{Jeong2}.     In the following all our numerical results  are for short--ranged potentials, unless stated otherwise.

We applied a Hubbard model to the interacting disordered ZGNRs and 
used a self-consistent HF approximation  (HFA); this is because the self-consistency provides an excellent approximation when both disorder and interactions are present \cite{Mac1,Mac2,Wang}. 
We include both electron-electron interactions and disorder in a tight-binding model at half-filling.
When $U=0$   disorder can be  treated exactly  in the HFA  while in the other limit, where disorder is absent, interaction effects can be  represented well by the HFA, which widely used  in graphene related systems \cite{Pis,Stau}.  Its results are consistent with those of 
density functional theory \cite{Yang}.   
The total Hamiltonian in the HFA is
\begin{eqnarray}
&&H=-\sum_{<ij>\sigma} t_{ij}c_{i\sigma}^{\dag}c_{j\sigma}+\sum_{i\sigma} \epsilon_{i}c_{i\sigma}^{\dag}c_{i\sigma}\nonumber\\
&+&U\sum_{i}  (n_{i\uparrow} \langle n_{i\downarrow}\rangle+\langle n_{i\uparrow}\rangle n_{i\downarrow}
-\langle n_{i\uparrow}\rangle \langle n_{i\downarrow}\rangle )\nonumber\\
&-&\frac{U}{2}\sum_i (n_{i\uparrow} +n_{i\downarrow} ),
\label{Ham}
\end{eqnarray}
where  $c_{i\sigma}^{\dag}$ and $n_{i\sigma}$ are the electron creation and  occupation operators at site $i$ with spin $\sigma$. Since the translational symmetry is broken, the Hamiltonian is written in the site representation.  In the hopping term the summation is over the nearest-neighbor sites (the average value of hopping parameters is $\langle t_{ij}\rangle=t\sim 3 eV$).
The eigenstates and eigenenergies  are computed numerically by  solving the tight-binding Hamiltonian matrix self-consistently.  
The self-consistent occupation numbers $\langle n_{i\sigma}\rangle$ in the Hamiltonian are
the sum of the probabilities to   find electrons of spin $\sigma$ at site $i$:
\begin{eqnarray}
\langle n_{i\sigma}\rangle=\sum_{E\leq E_F} |\psi_{i\sigma} (E) |^2 .
\label{occnum}
\end{eqnarray}
The sum is over  the occupied eigenstates with energy $E$ below the Fermi energy $E_F$. 
Note that  $\{  \psi_{i\sigma}(E)  \}$ 
represents an eigenvector of the tight-binding Hamiltonian matrix with  energy $E$.   For notational simplicity, we suppress  its dependence on $E$ from now on.    
The ratio between the disorder strength and interaction strength is $\kappa=\Gamma\sqrt{n_{imp}}/U\ll 1$ in the weak disorder regime.    Varying the strength  $\Gamma$ is approximately equivalent to changing $\sqrt{n_{imp}}$.
In this work, the ribbon width was set to $w = 7.1\,\AA$ 
and the on-site repulsion was $U=t$.   To investigate very long ZGNRs  it is vital to use sparse matrix diagonalization techniques.

\section{Quantized fractional charge of midgap state}

For the sake of  clarity we briefly summarize the results  we obtained in Ref. \cite{Jeong2}.  The midgap states with energy $|E|\approx \Delta_s/2\ll \Delta/2$ and edge charge $q_A\approx 1/2$ represent {\it soliton} states.  
 They consist of almost {\it equal} contributions from the valence $R$ and conduction band $L$ states (or from the valence band $L$ and conduction band $R$ states) with energies near $-\Delta/2$ and $\Delta/2$, respectively, as shown in Fig.~\ref{degGap}. 
A soliton state has small disorder induced charge fluctuations.   In addition,   spin-split states are also present  \cite{Jeong1}, as in  a Mott-Anderson insulator  \cite{Dob}.
For a given disorder realization, greater spin-splitting occurs for 
states with $|E|\approx \Delta_s/2$ than for those with
$|E|\approx \Delta/2$.  
In the limit where disorder strength $\Gamma\rightarrow 0$ and  ribbon length $\ell \rightarrow \infty$ the energy of a soliton decreases   toward $E=0$ and $q_A\rightarrow 1/2$
with   very small fluctuations, i.e.,the value of the  fractional charge approaches $e/2$.

\begin {figure}[!hbpt]
\begin{center}
\includegraphics[width=0.35\textwidth]{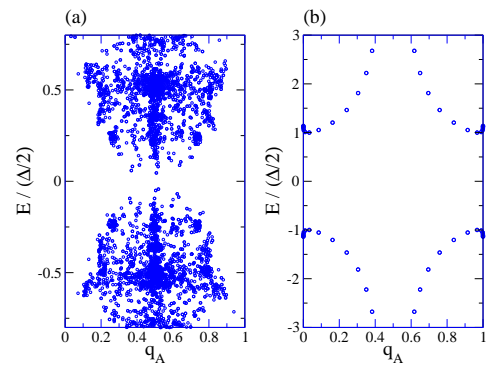}
\caption{ (a) Plot of $q_A$ for disorder potential strength $\Gamma=0.1t$, where each point represents the probability of finding an electron of a gap-edge state on $A$ carbon sites, $q_A$. The disorder realization number is $N_D=300$, and the ribbon length is $\ell=196.8 $\AA. Here, the impurity or defect--to--carbon atom ratio is $n_{imp}=0.1$. 
The gap size is $\Delta_s\approx 0.12\frac{\Delta}{2}$.   (b) Plot of $q_A$ for $U=t$ in the absence of disorder, where the on-site electron repulsion and hopping parameter are indicated by $U$ and $t$, respectively. }
\label{fluc}
\end{center}
\end{figure}

\begin {figure}[!hbpt]
\begin{center}
\includegraphics[width=0.4\textwidth]{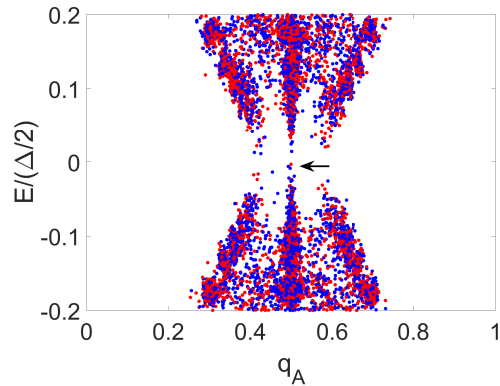}
\caption{ Plot of $q_A$ for  $\Gamma=0.03t$,  $U=t$, $n_{imp}=0.1$,
$l=1232.5$\AA, and $N_D=2166$.    Zero energy states with $q_A$ rather close to $1/2$ are indicated by an arrow.  Blue (red) dots are for spin-up (-down) states.}
\label{weak}
\end{center}
\end{figure}

Here we provide a new and  efficient  way to analyze  numerical results.  The numerical results are presented in the following  way:   For each HF quasiparticle state with energy $E$  and spin $\sigma$ we compute  the total probability density on A carbon sites, denoted by 
\begin{equation}
q_A=\sum_{i\in \:A}|\psi_{i\sigma}(E)|^2.    
\end{equation}
We plot all the possible values of $(E,q_A)$. This  plot makes it easier 
to delineate physics behind charge fractionalization.
We find that disorder behaves similar to a {\it singular} perturbation on zigzag edge electronic states \cite{zero}. This singular perturbation is  analogous to the nonperturbative coupling between the left and right wells of a double quantum well (the  nonperturbative aspect can be seen by  using   instantons of the inverted double well potential  \cite{singular}).   
A disorder potential or a magnetic field can produce {\it drastic} changes in the electron wave functions  \cite{Kim,com2}, see Figs.~\ref{fluc}--(a) and \ref{fluc}(b).  
Figure ~\ref{fluc}--(a) shows the distribution of $(E,q_A)$ for $\Gamma=0.1t$.   Note that particle-hole symmetry (chiral symmetry) is {\it broken}.   
Even a weaker disorder potential with $\Gamma=0.03t $ produces similar 
drastic changes in the energy spectrum when compared to the disorder-free behavior, see Fig. \ref{weak}.  Note that in this weak disorder regime  charge fractionalization is more accurate:  Midgap states shown in the figure has $q_A$  very close to $1/2$.
In contrast to the case 
of $\Gamma=0$,  shown in Fig.~\ref{fluc} (b),  there are numerous states with $q_A\approx 1/2$ in the energy range $|E|<\Delta/2$.    
If the disorder potential experienced by the left and right edges differs, then charge  values will deviate from  $1/2$.

\begin {figure}[!hbpt]
\begin{center}
	\includegraphics[width=0.4\textwidth]{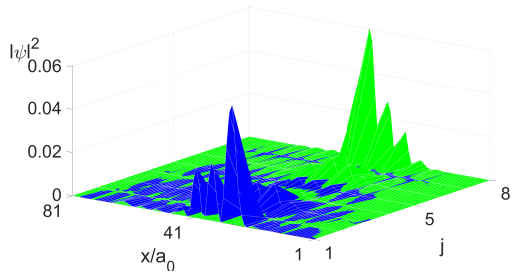}
	\caption{  Plot of the probability density of a fractionalized gap-edge state in the presence of  off--diagonal disorder.  It is a mixed chiral state; green and blue represent different chiralities.   Its energy is $E=-0.019t$ with
	$q_A=0.495$.  The range of hopping parameters is $0.94 <t_{ij}/t  < 1.06$ and the ribbon length is $\ell=199.3$\AA.  A  zigzag ribbon consists of zigzag lines.
In this ribbon they  are labeled from $j=1$ to $8$.}
	\label{fraccharge1}
\end{center}
\end{figure}

\begin {figure}[!hbpt]
\begin{center}
	\includegraphics[width=0.4\textwidth]{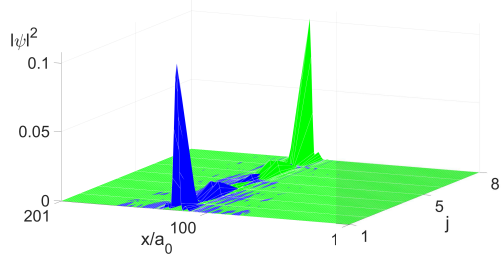}
	\caption{ Same as in Fig.~\ref{fraccharge1} but with a longer ribbon length $\ell=494.5 $\AA.  The gap-edge-state  energy is $E=0.055t$ with
	$q_A=0.51$.  The range of hopping parameters is $0.94 <t_{ij}/t  < 1.06$.  }
	\label{fraccharge2}
\end{center}
\end{figure}

We have also examined off-diagonal disorder.  In the presence of off-diagonal disorder, a zigzag edge site is connected  to two other carbon atoms while a site away from the edges are connected to three other carbon atoms, but the hopping parameter is not the same for every site.  Despite this our numerical results show that
fractional midgap states do  exist as in diagonal disorder, see Figs.~\ref{fraccharge1} and \ref{fraccharge2}.
 The  network topology, i.e., how many carbon atoms each site of the zigzag edges is connected to, is crucial   for charge fractionalization.  Whether disorder  is off--diagonal or diagonal is immaterial.

We find that the localization length along the edges decreases as $|E|$ decreases toward $\Delta_s/2$.  A small localization length   means that the repulsive energy between an electron in a soliton state and an  added  electron in another  soliton state 
can be small since they can avoid each other.  This effect determines the magnitude of $\Delta_s$.

\section{Linear tunneling density states}

\begin {figure}[!hbpt]
\begin{center}
\includegraphics[width=0.3\textwidth]{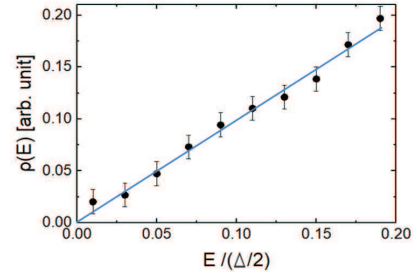}
\caption{ Plot of DOS $\rho(E)$ for disorder potential strength $\Gamma=0.18t$.
Here, the disorder realization number $N_D=4409$, the impurity or defect-to-carbon atom ratio $n_{imp}=0.1$, and ribbon length $\ell=740.46$ \AA. 
The histogram interval is $\delta E=0.02\frac{\Delta}{2}$. The solid line represents a
linear fit to $\rho(E)$. }
\label{dosplot2}
\end{center}
\end{figure}

The $e/2$ fractional charge fluctuations decrease as $|E|\rightarrow \Delta_s/2$. 
Thus, we investigated the effect of this  behavior on the DOS near the Fermi energy.  (Note that the tunneling DOS measures the number of quasiparticle excitations of the interacting disordered ZGNR).
We  examined  longer ZGNRs than those  in Ref. \cite{Jeong2}.   This allows us to extract the behavior of the DOS in the limit $E\rightarrow 0$.  We performed finite-size calculations and computed the DOS given by 
\begin{equation}
\rho(E)=\frac{D_{\delta E}(E)}{l N_D \delta E}, 
\end{equation}
where $D_{\delta E}(E)$ is the total 
number of states in the energy histogram interval $ \delta E$ and $N_D$ is the number of disorder realizations.
We defined the critical point $\Gamma_c$ as the value where  $\Delta_s$ is zero, i.e., where the gap closes. The heuristic argument given above suggests that the DOS at $\Gamma_c$ is linear near the Fermi energy.  
The DOS result for $\Gamma=0.18t\gtrsim \Gamma_c$  is plotted in Fig.~\ref{dosplot2}.   
Our numerical results show that  the energy range where the  DOS is linear  increases as ribbon length $\ell$ grows and that fluctuations in the DOS also decreases.  
Note that  $\Gamma_c $ does {\it not} represent a metal--insulator transition point, and the gap-edge states are all localized
in the interacting disordered ZGNR \cite{com1}.  Note also that $\Gamma_c $ decreases as $\ell$ increases  (this is a finite-size effect).

\begin {figure}[!hbpt]
\begin{center}
\includegraphics[width=0.3\textwidth]{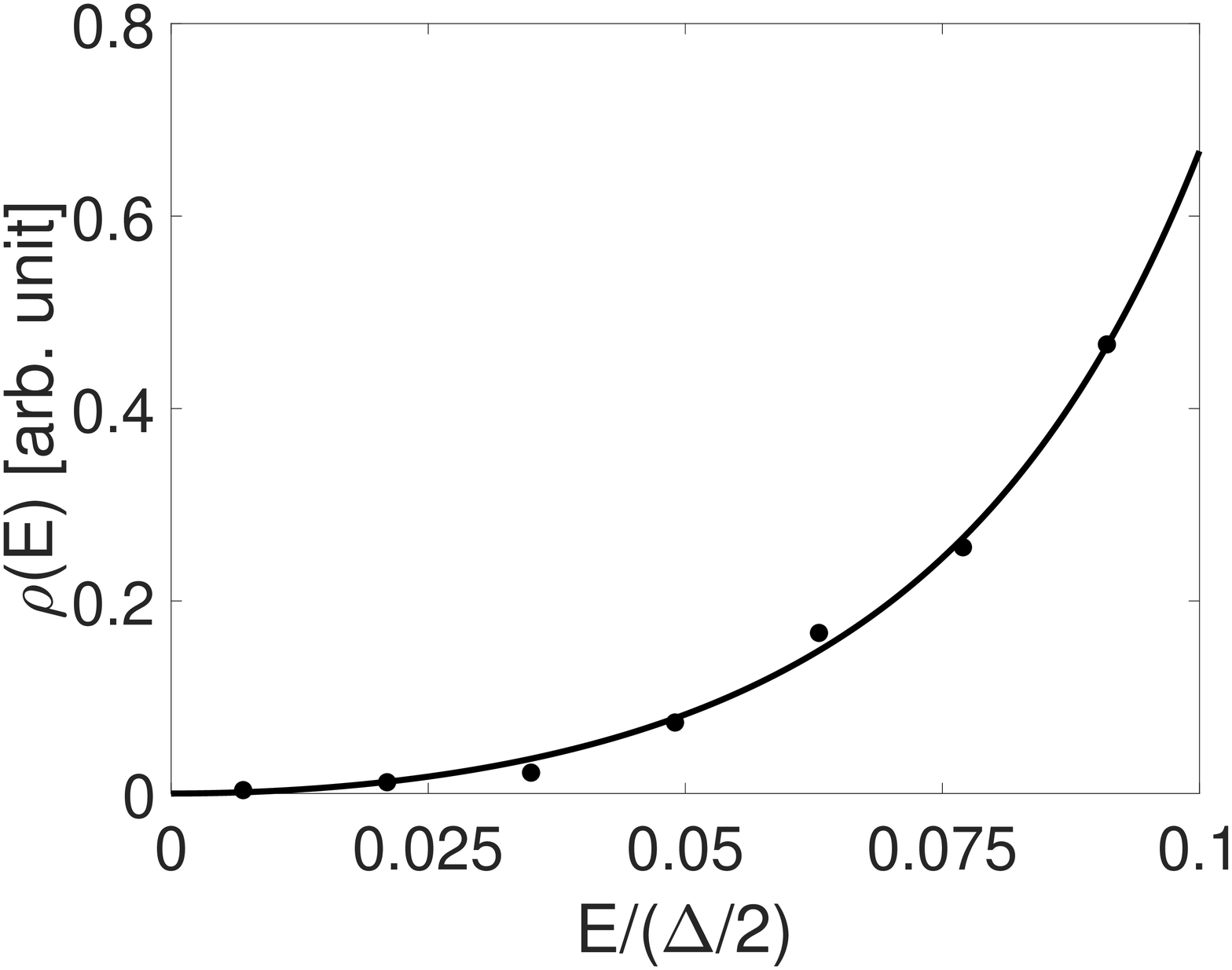}
\caption{  The DOS displays an exponential soft gap near $E=0$. Parameters are   $U=t$,
$\Gamma=0.03t$, $n_{imp}=0.1$,
$l=1232.5$\AA, and $N_D=2166$.  Histogram interval is $0.014\frac{\Delta}{2}$.}
\label{prof}
\end{center}
\end{figure}

In the limit $\ell\rightarrow \infty$   our results for $\Gamma< \Gamma_c$ suggest that the DOS decreases exponentially
to zero  as $E\rightarrow 0$, see Fig.~\ref{prof}.    The  shape of the resulting soft gap can be fitted well with an exponential form  of
\begin{equation}
\rho(E)=A(e^{\alpha x^2}-1), 
\end{equation}
where $x=E/(\Delta/2)$ and the fitting parameters are $A=0.164$ and  $\alpha=162$.  This suggests that the size of the exponential gap is of the order of  $\sim 0.05\frac{\Delta}{2}$.

\section{Spin--charge separation}

\begin {figure}[!hbpt]
\begin{center}
\includegraphics[width=0.4\textwidth]{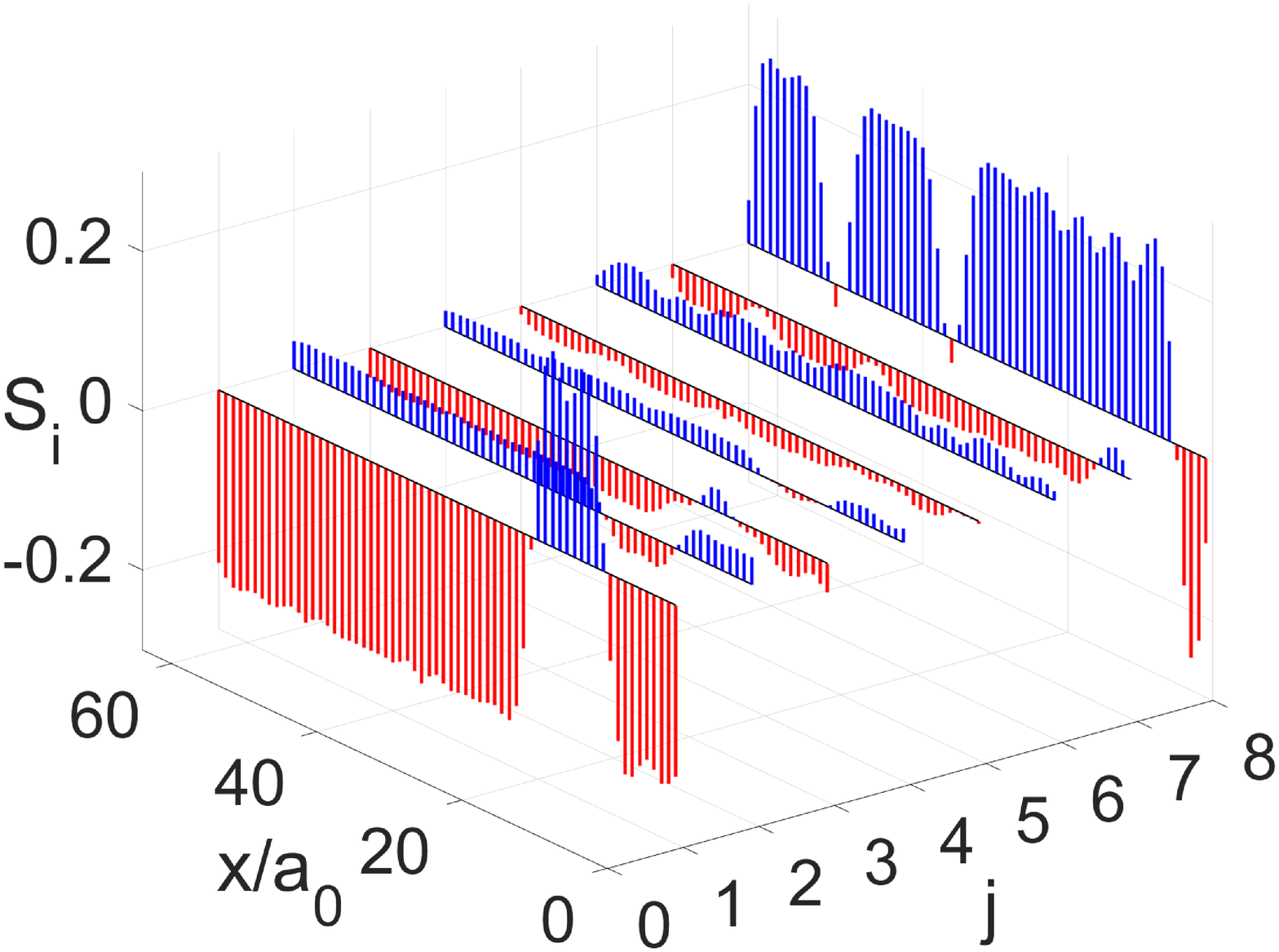}
\caption{Net site spin values $S_{i}=S_{i\uparrow}+S_{i\downarrow}=n_{i\uparrow}-n_{i\downarrow}$ are plotted, where $n_{i\sigma}$ is the site occupation number for spin $\sigma$.  A  zigzag ribbon consists of zigzag lines.
In this ribbon they  are labeled from $j=1$ to $8$.  Blue (red) lines indicate positive (negative) spin values. This result is for one disorder realization with diagonal disorder. The parameters  are $U=t$. }
\label{disorderanti}
\end{center}
\end{figure}

An interacting disordered ZGNR displays  antiferromagnetism that is weakly perturbed, see Fig.~\ref{disorderanti}.  As shown in Fig.~\ref{degGap}--(a) a zigzag ribbon consists of zigzag lines.  Away from the outer two zigzag edges a
zigzag line  inside the ribbon is mostly antiferromagnetically coupled  with the neighboring  two zigzag lines.  Magnetization is mostly ferromagnetic  on each  of the two boundary zigzag edges but the two zigzag edges are antiferromagnetically coupled.  On the left zigzag edge the site spin direction flips in a region.  Also note that on the right zigzag edge the site spin values are nearly zero in two regions. These effects are due to the singular nature of disorder, as we explain below.

\begin {figure}[!hbpt]
\begin{center}
\includegraphics[width=0.3\textwidth]{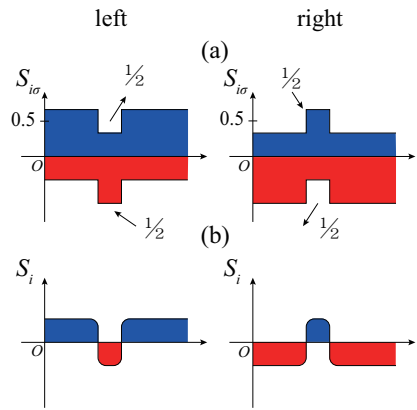}
\caption{(a) Site-spin $z$ components, $S_{i\uparrow}=n_{i\uparrow}$ (blue) and $S_{i\downarrow}=-n_{i\downarrow}$ (red), plotted along edges.  Here, $n_{i\sigma}$ is the site occupation number. The left and right figures correspond to the left and right zigzag edges, respectively. The number $1/2$ indicates a removed or added electron occupation number.   (b) Net site spin values $S_{i}=S_{i\uparrow}+S_{i\downarrow}$ are plotted. }
\label{profa}
\end{center}
\end{figure}

\begin {figure}[!hbpt]
\begin{center}
\includegraphics[width=0.3\textwidth]{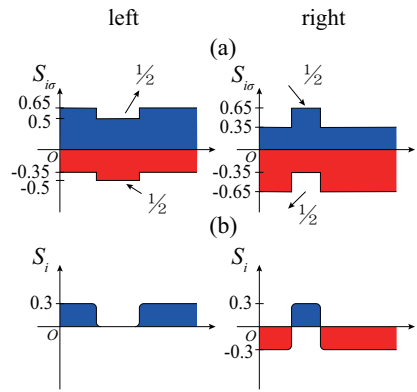}
\caption{ Plot of total $z$ component of ground state site spin $S_{i}$. Process of spin-charge separation is displayed in  (a) and  (b).}
\label{profb}
\end{center}
\end{figure}

Let us introduce disorder into the pure SPT phase and
try to understand how the singular  disorder potential disrupts  the SPT phase.   Suppose that the site occupation numbers of the disorder-free left edge are $n_{i\uparrow}=0.7$ and $n_{i\downarrow}=0.3$. Then, those of the right edge are $n_{i\uparrow}=0.3$ and $n_{i\downarrow}=0.7$, respectively. 
Assume that disorder generates one spin-up and one spin-down occupied soliton state near the gap edge displaying charge fractionalization. In other words, a spin-up electron on the left zigzag edge of the interacting ZGNR is converted into  two $e/2$ fractional charges, one of which resides on the left zigzag edge while the other resides on the right zigzag edge [Fig.~\ref{profa} (a)]. Similarly, a spin-down electron on the right zigzag edge is also replaced by two $e/2$ fractional charges, with one each residing on the left and right zigzag edges [Fig.~\ref{profa} (a)]. Hence, the total z component of the site spin $S_i=S_{i\uparrow}+S_{i\downarrow}$ on the zigzag edges changes sign along the edge direction, as shown in Fig.~\ref{profa} (b). The total occupation number of each site $n_i=n_{i\uparrow}+n_{i\downarrow}$ is now close to one (i.e., the ZGNR is half-filled). 
Note that the   disorder potential  creates  an  {\it even number of   solitons}    to minimize the energy cost of double occupancy 
of a site (a soliton consists a pair of fractional charges).  Thus, even if the disorder potential is  weak it can  still disrupt   the SPT  phase.  
In addition, the magnetic zigzag edge reconstruction can also lead to a  {\it spin--charge separation} \cite{Su,GY,Lee}.   
Figures \ref{profb} (a) and \ref{profb} (b) show how a charge  fractionalization process results in an object $  (e_L,0)$ that displays spin--charge separation.
Here  $e_L$  denotes an electron charge located on the left  edge and number  $0$ means no spin.  When such an object moves along the zigzag edge it will carry charge but {\it no} spin.

\section{Main physics}

Disorder has profound effects on a ZGNR as it breaks particle--hole symmetry (chiral symmetry),  in addition to  inducing spin splitting.  Other symmetries are also broken:  translational, reflection, rotational, and  inversion symmetries.  Spin-rotational symmetry is  spontaneously broken \cite{broken}.  Time-reversal symmetry is already broken by antiferromagnetism. 
There is no symmetry that will protect edge states in 
an interacting disordered  ZGNR.  Moreover,  as we already remarked, there are doubly degenerate   ground states that  are  connected to each other via reversal of electron spin directions (see Fig.~\ref{degGap}).  All this suggests that an interacting disordered ZGNR  is qualitatively different from the disorder-free  interacting  ZGNR.

We now explain the essential physics of charge fractionalization and the physical  nature of interacting disordered ZGNRs.  The e/2 charges are a result of the subtle interplay between network topology of the underlying lattice, electron correlation and disorder.  In each disorder realization particle--hole symmetry (chiral symmetry) is broken, but  after disorder averaging the symmetry is approximately restored.  This implies that the {\it average} edge charge is $e/2$ at each energy $|E| <\Delta/2$, but   with a significant  charge variance.  However,  if  the tunneling DOS develops a soft gap  \cite{Efros,Mac1}, then the charge variance near zero energy will be negligible in the weak disorder regime, see Fig.~\ref{weak}.
What is the physical origin of a soft gap?  The essential physics is that it is difficult for the tunneling electron  to avoid other electrons since it takes long time for interacting electrons to diffuse away from each other (see Girvin and Yang, Ref. \cite{GY}, pp. 290 and 645).  Our numerical simulation shows that an interacting disordered ZGNR cannot be reached iteratively from a disorder--free chiral  SPT state.
Moreover, an interacting disordered ZGNR has a doubly degenerate ground state, $e/2$ fractional charges, spin--charge separation, and broken chiral symmetry.     Thus we expect that it is  in  a topological ordered phase 
rather than in an SPT   phase (see Wen \cite{Wen1} for the distinction between them).  An interacting disordered ZGNR is somewhat analogous to topologically ordered Laughlin states.  In both systems
fractional charge and  ground state degeneracy are intimately related \cite{Wen1,GY}.

\section{Conclusions}

In conclusion, an interacting disordered ZGNR is a one-dimensional topologically ordered  insulator with $e/2$ solitonic fractional charges and with two-fold ground state degeneracy.  Even a weak disorder potential behaves similar to a singular perturbation, producing 
spin-splitting and drastically modifying the energy spectrum.
We conducted a numerical study showing that the DOS is linear at the critical disorder strength. Measurement of the $I\mbox{-}V$ curve may thus provide evidence for the presence of fractional charges in an interacting disordered ZGNR. We also found that  spin--charge separation and fractional edge charges play a significant role in the reduction of edge antiferromagnetism.
We hope that our work will stimulate experimental tests investigating the presence of $e/2$ fractional charges in interacting disordered ZGNRs. However, several experimental possibilities and challenges exist.
In particular, investigation of tunneling between zigzag edges, as in fractional quantum Hall bar systems \cite {Kang}, may be fruitful. Quantum shot noise may directly measure \cite{de} the tunneling fractional charge of a ZGNR. Resonant tunneling measurement through a quantum dot structure made of a rectangular ZGNR may also be explored \cite{Gold}.  
Finally, it would  be  interesting to investigate    other  zigzag nanoribbon systems that exhibit antiferromagnetism, e.g., silicene and boron 
nitride nanoribbons \cite{Yao,Bar}.  Disorder can couple the left and right zigzag edges and  lead to  charge  fractionalization.

\section*{Acknowledgments}
This research was supported by the Basic Science Research Program
through the National Research Foundation of Korea (NRF), funded by the
Ministry of Education, ICT, and Future Planning (MSIP) [NRF-2018R1D1A1A09082332 (S.R.E.Y.) and NRF-2019R1F1A1062704 (M.C.C.)].

\appendix

\section{Particle  fractionalization  in other  systems}

Kitaev's chain \cite{Kit}, polyacetylene, and interacting disordered ZGNRs  all  have end states.   There are similarities and differences between them \cite{yang1}.  It is both interesting and instructive to take note of them.

\begin {figure}[!hbpt]
\begin{center}
\includegraphics[width=0.25\textwidth]{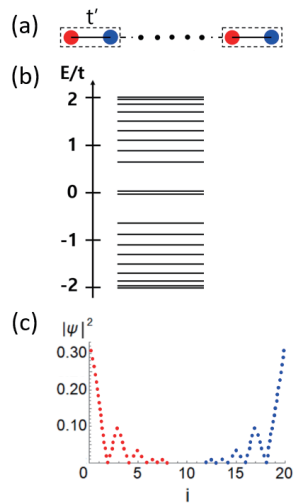}
\caption{(a)  Finite-length dimer chain with unit cell containing two carbon atoms connected by  single bond.   The intra cell hopping  $t'$ is smaller than the inter cell hopping $t$.  (b) Tight-binding energy spectrum.   Two nearly  degenerate gap states  exist.  (c) Probability density of  a gap state as  function of site index $i$.   A peak is apparent at the red (blue) site at the left (right) end. The probability densities of the bonding and antibonding states are almost identical.}
\label{poly}
\end{center}
\end{figure}

\begin {figure}[!hbpt]
\begin{center}
\includegraphics[width=0.4\textwidth]{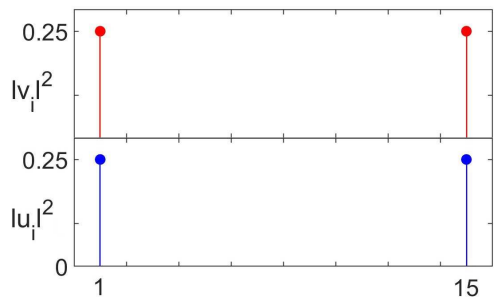}
\caption{Kitaev's chain has two degenerate zero energy states.  They can be combined to give one bonding and one antibonding states, see Fig.~\ref{endsts}.  For each of these  states the probability to find an electron  (hole)  at  site $i=1,\ldots, 15$ is $|u_i|^2$  ($|v_i|^2$).  
We have set the hopping parameter equal to the value of the gap $t=\Delta$ and  the chemical potential $\mu=0$.   A Majorana zero mode (half a real fermion mode) at each end of the chain is displayed. }
\label{zeroendmode}
\end{center}
\end{figure}

Chiral symmetry  guarantees existence  of edge states in  ZGNRs and polyacetylene.   
However, in the presence of disorder there is an important difference between ZGNRs and polyacetylene.    
Let us consider the Su-Schrieffer-Heeger effect in detail \cite{SSH}.   Consider  finite-length  polyacetylene in one of the dimerized phases, see Fig.~\ref{poly}.  The electron density is uniform with occupation number   $n_i=1$ at all  sites $i$.    Here    two nearly  degenerate soliton end  states  exist (bonding and antibonding states); one will be occupied and the other unoccupied.  Disorder will spilt these states because  their wave functions  are somewhat different.  Then two possibilities are present.  (a) The boundary fractional charges will suffer quantum charge fluctuations  because the energy splitting is small (only a large energy splitting will suppress quantum charge fluctuations, see an insightful discussion by Girvin \cite{Girvin}).  (b) Disorder will most likely not couple the left and right ends equally.  Hence the boundary charge will not be exactly $e/2$.  There is as yet no conclusive experimental evidence for fractional charges in polyacetylene (but spin-charge separation was observed).

Consider Kitaev's toy model of one-dimensional {\it p}-wave superconductivity, which has relevance to topological superconductors \cite{Bennakker}.   Particle--hole symmetry plus bulk edge correspondence guarantees   the presence of   zero-energy modes.  
It exhibits  a  charge {\it neutral} particle that is  divided between the two ends of the chain.  These Majorana zero modes are expected to display non-Abelian statistics \cite{NonAbel}.  Figure~\ref{zeroendmode} displays such zero modes  of a finite length chain. (This is a  well known result and we show it here just for comparison with polyacetylene and ZGNRs).

Now let us discuss edge states of a ZGNR.  The probability density of the midgap states is fractionalized equally between the left and right zigzag edges \cite{Jeong0}.   It is similar to fractionalization occurring at the end points of  polyacetylene \cite{Su} and   Kitaev's chain \cite{Kit}.   However, in  interacting disordered ZGNRs fractional charges reside on the zigzag edges that form  the  {\it side}  boundary of the ribbon, see Fig.~\ref{fracchar}.   In addition, the presence of a gap $\Delta_s$ protects the fractional charges against quantum charge fluctuations.


\begin{thebibliography}{9}
\bibitem{compoly} A good example of an SPT phase is polyacetylene.  Chiral symmetry guarantees  that end states exist.
\bibitem{dimension} A ZGNR is a quasi--one--dimensional system and not one--dimensional.  A ribbon obviously has a width.  Often properties of a  quasi--one--dimensional system may be different from those of a strictly one-dimensional system.
\bibitem{Insul} In an Anderson insulator  disorder induces localization of electron states.  In a	 
	 Mott     insulator the Fermi level is in the 
	 gap induced by many-body interactions.  In a Mott-Anderson insulator both disorder and many-body 
	 interactions are important.  Localized states often display spin--splitting.
\bibitem{topoph} In an  SPT phase symmetry determines  properties of the system while in a 
	topological ordered phase properties are beyond symmetry description.
\bibitem{zero} The number of zero modes in a ZGNR is proportional to the ribbon length.   Zero modes are  important in topological insulators because their presence may be related to  ground--state degeneracy.
\bibitem{singular} An instanton is a non-perturbative effect.   When it is  applied to a double quantum well one finds that the energy splitting between the bonding and 
	antibonding states is $\Delta E\propto e^{-1/\lambda}$, which is singular in the coupling constant $\lambda$.  In field theory an instanton is used to  connect two vacua.
\bibitem{com1} 
	The gap of  a disordered ZGNR is  filled with localized  states, whereas     the states outside the gap are  delocalized  \cite{Jeong2}.    Thus, the usual one-dimensional localization theory does not apply to the Dirac electrons of GNRs  \cite{Ando,Neto}. 
  A short-ranged disorder potential  induces stronger localization of gap--edge states along the zigzag edges  than a long-ranged disorder potential  \cite{Lima}.
\bibitem{broken} Spontaneously broken symmetry means:  the Hamiltonian is invariant under spin rotation, but the ground state  is not invariant.
\bibitem{NonAbel}  	A non-Abelian particle is neither a fermion nor a boson. On	exchange of two non-Abelian particles their total wave function acquires a matrix Berry phase
\end{thebibliography}
\end{document}